\documentclass[%
 reprint,
 amsmath,amssymb,
 aps,
]{revtex4-2}
\usepackage{braket}
\usepackage{here}
\usepackage[dvipdfmx]{graphicx}
\usepackage{dcolumn}
\usepackage{bm}
\usepackage{newtxtext,newtxmath}
\usepackage{hyperref}
\hypersetup{
setpagesize=false,
bookmarksnumbered=false,%
bookmarksopen=false,%
colorlinks=true,%
linkcolor=blue,
citecolor=blue,
urlcolor=blue,
}
\usepackage[normalem]{ulem}
\usepackage{xcolor}
\usepackage{lineno}
\newcommand*\patchAmsMathEnvironmentForLineno[1]{%
  \expandafter\let\csname old#1\expandafter\endcsname\csname #1\endcsname
  \expandafter\let\csname oldend#1\expandafter\endcsname\csname end#1\endcsname
  \renewenvironment{#1}%
     {\linenomath\csname old#1\endcsname}%
     {\csname oldend#1\endcsname\endlinenomath}}%
\newcommand*\patchBothAmsMathEnvironmentsForLineno[1]{%
  \patchAmsMathEnvironmentForLineno{#1}%
  \patchAmsMathEnvironmentForLineno{#1*}}%
\AtBeginDocument{%
\patchBothAmsMathEnvironmentsForLineno{equation}%
\patchBothAmsMathEnvironmentsForLineno{align}%
\patchBothAmsMathEnvironmentsForLineno{flalign}%
\patchBothAmsMathEnvironmentsForLineno{alignat}%
\patchBothAmsMathEnvironmentsForLineno{gather}%
\patchBothAmsMathEnvironmentsForLineno{multline}%
}
\newcommand{\bs}{\boldsymbol}

\newcommand{\bQ}{\mathbf{Q}}
\newcommand{\bE}{\mathbf{E}}
\newcommand{\bH}{\mathbf{H}}
\newcommand{\bI}{\mathbf{I}}
\newcommand{\be}{\mathbf{e}}
\newcommand{\bk}{\boldsymbol{k}}
\newcommand{\bn}{\mathbf{n}}
\newcommand{\br}{\mathbf{r}}
\newcommand{\bv}{\boldsymbol{v}}
\newcommand{\bkhat}{\hat{\boldsymbol{k}}}
\newcommand{\sigmae}{\bs{\upsigma}^\mathrm{e}}
\newcommand{\sigmaa}{\bs{\upsigma}^\mathrm{a}}

\newcommand{\Add}[1]{{#1}}
\newcommand{\Bdd}[1]{{#1}}
\newcommand{\Delete}[1]{}
\newcommand{\Comment}[1]{}

\begin{document}

\title{Active nematic liquid crystals under a quenched random field}

\author{Yutaka Kinoshita}
\author{Nariya Uchida} 
\email{nariya.uchida@tohoku.ac.jp}
\affiliation{%
 Department of Physics, Tohoku University, Sendai 980-8578, Japan
}%

\date{\today}

\begin{abstract}
Coupling between flow and orientation is a central issue in understanding the collective dynamics of active biofilaments and cells. Active \Bdd{stresses} generated by motor activity destroy (quasi-)long-range orientational order and induce chaotic \Bdd{flows with many vortices}. 
In \Delete{the} cellular and subcellular environment, alignment is also hindered by heterogeneous filamentous structures in \Delete{the} extracellular matrix and various \Bdd{intracellular organelles}. 
Here we address the effects of a quenched random field on the flow patterns and orientational order in two-dimensional active nematic liquid crystals. We \Bdd{find} that the director dynamics \Bdd{becomes} frozen above a critical disorder strength. For sufficiently strong randomness, the orientational correlation function decays exponentially with \Delete{the} distance, reproducing the behavior of passive random-field nematics. In contrast, the flow velocity decreases only gradually \Bdd{with increasing disorder}, and \Bdd{exhibits} a logarithmic spatial correlation \Bdd{under strong randomness}. \Bdd{We identify} the threshold between the activity- and disorder-dominated regimes  
\Bdd{and examine} its dependence on the activity parameter.
\end{abstract}


\maketitle


{\it \label{sec::intro}Introduction.} --
Collective motion of cells and biofilaments \Bdd{is} of vital importance to life at various stages, such as cell division, morphogenesis, cell migration, and apoptosis. The dynamics \Bdd{are} driven by molecular motors and facilitated by orientational ordering of active elements \Bdd{with} slender shapes. While apolar interaction induces nematic order, active stresses generated by motor activity destroy (quasi-)long-range orientational order and generate chaotic flows with many vortices, which are known as active turbulence~\cite{wensink2012meso,doostmohammadi2018active, alert2022active}. Active nematic turbulence \Bdd{has been} demonstrated in a two-dimensional suspension of microtubules and kinesin-motor-complexes~\cite{sanchez2012spontaneous}. 
Topological defects are also found in colonies of cells~\cite{doostmohammadi2016defect,saw2017topological,kawaguchi2017topological} and 
\Bdd{multi}cellular organisms~\cite{maroudas2021topological}, 
and their biological functions have been revealed.

The flow patterns of active nematics can be controlled by friction with the substrate~\cite{guillamat2016control,guillamat2016probing,guillamat2017taming, thampi2014active,srivastava2016negative,thijssen2020active}, external fields~\cite{green2017geometry,krajnik2020spectral,kinoshita2023flow}, and confinement~\cite{giomi2012banding,doostmohammadi2017onset,shendruk2017dancing,norton2018insensitivity,coelho2019active, rorai2021active}. 
\Bdd{Theoretical studies have examined the effects of uniform external fields, such as in}
a model of an active pump using a Frederiks twisted cell~\cite{green2017geometry}. 
A three-dimensional simulation of active nematics under an electric field found 
a direct transition from \Bdd{active} turbulence to a uniformly aligned state~\cite{krajnik2020spectral}, while a laning state intervenes in two dimensions~\cite{kinoshita2023flow}.
Laning states are also obtained in numerical simulations with isotropic~\cite{thampi2014active,srivastava2016negative} or anisotropic~\cite{thijssen2020active} friction, and 
\Bdd{have also been explored} experimentally~\cite{guillamat2016control}. 
Confinement also results in various director patterns such as the laning state and vortex lattice~\cite{doostmohammadi2017onset,norton2018insensitivity,rorai2021active, coelho2019active,shendruk2017dancing}. Recently, attention has \Bdd{shifted toward} 
couplings of active nematics with non-uniform fields, such as friction from micropatterned surfaces~\cite{thijssen2021submersed}, curvature of epithelial tissues~\cite{vafa2022active}, spatially varying activity~\cite{ronning2023defect,partovifard2024controlling}, and composition~\cite{assante2023active}.

On the other hand, the behavior of active nematics in a randomly heterogeneous environment 
is an open issue. 
Cells in tissues are in contact with the extracellular matrix \Bdd{which contains} 
fibrous material such as \Bdd{collagen and cellulose}. 
Collagen fibers form anisotropic networks that guide the migration of cells~\cite{thrivikraman2021cell}. 
Plant cellulose is also used as a scaffold for in vitro culture of neural stem cells~\cite{couvrette2023plant}. 
Microtubules in cells are entangled with other components of the cytoskeletal network such as actins 
and intermediate filaments, which hinder alignment. 
The cytoplasm also contains a number of proteins that cyclically change their shapes and generate random hydrodynamic forces. They not only enhance diffusion in the cell~\cite{mikhailov2015hydrodynamic}, 
but may also contribute to the disorientation of active cytoskeletal filaments. 
To elucidate the effects of heterogeneous anisotropic environments on active nematic flows, 
we address the effects of quenched random fields in this paper.

%
%
The effect of a quenched random field on nematic liquid crystals has been studied in a model of nematic elastomers~\cite{yu1998exponential}. The numerical study showed that the orientational correlation function decays exponentially as a function of \Delete{the} distance, and that the correlation length also decays exponentially with the disorder strength.

Recently, the effects of a random field on dry and wet active nematics have been addressed in Refs.~\cite{kumar2020active, kumar2022active}. In both cases, the quenched disorder slows the dynamics of topological defects. The orientational correlation 
\Bdd{function exhibits} a crossover from algebraic to exponential decay \Bdd{for} dry active nematics, 
\Bdd{whereas} wet active nematics \Bdd{display} a slow coarsening behavior. 
However,
\Bdd{the} flow \Bdd{behavior of} active nematics \Bdd{under} quenched randomness remains \Bdd{largely} unexplored, despite its potential importance in cytoskeletal dynamics and the collective migration of cell colonies.

In the present paper, we focus on the flow \Bdd{behavior} and show that its response to quenched randomness is in marked contrast with that of the orientational order. We also vary the randomness over a wide range and identify a transition from the weak to strong disorder regimes for wet active nematics.
\\


{\it \label{sec::model}Model.} --
The orientational order of a two-dimensional nematic liquid crystal
is described by the symmetric and traceless 
tensor $Q_{ij}=S\left(n_i n_j - \frac12 \delta_{ij} \right)$, 
where $S$ is the scalar order parameter 
and $\bs{n}=\left(\cos\theta,\sin\theta\right)$ is the director.
The dynamical equations of active nematics 
in the dimensionless form read~\cite{doostmohammadi2018active} 
\begin{equation}
\left(\partial_t+\boldsymbol{v}\cdot\boldsymbol{\nabla}\right)\bs{v}
=
\frac{1}{\mathrm{Re}}{\nabla^2} \bs{v}-\bs{\nabla}p+\bs{\nabla}\cdot\bs{\upsigma}
\label{eq:dvdt}
\end{equation}
and
\begin{equation}
\left(\partial_t+\boldsymbol{v}\cdot\boldsymbol{\nabla}\right)\mathbf{Q}
=
\lambda S\mathbf{u}+\mathbf{Q}\cdot\bs{\upomega}-\bs{\upomega}\cdot\mathbf{Q}+\gamma^{-1}\mathbf{H}.
\label{eq:dQdt}
\end{equation}
Here, $\bs{v}$ is the normalized flow velocity\Bdd{,} 
which satisfies the incompressibility condition
$\bs{\nabla}\cdot\bs{v}=0$,
$p$ is the pressure and 
$\bs{\upsigma}$ is the stress tensor. 
The flow properties are characterized by
the Reynolds number  $\mathrm{Re}$, 
the flow alignment parameter $\lambda$, 
and the rotational viscosity $\gamma$,
and 
$u_{ij}=(\partial_i v_j+\partial_j v_i)/2$ and $\upomega_{ij}=(\partial_i v_j-\partial_j v_i)/2$ 
are the symmetric and antisymmetric parts of velocity gradient tensor, respectively.
\Bdd{We hereafter refer to} $\omega=\omega_{xy}$ \Bdd{as} the vorticity.
{We assume $0<\lambda <1$; 
a positive value of $\lambda$
corresponds to elongated or rod-like elements and 
$|\lambda|<1$ \Bdd{places the system in} 
the flow-tumbling regime\Bdd{,} where 
no stable director orientation exists in a uniform 
shear flow~\cite{de1993physics, edwards2009spontaneous}.
The molecular field $H_{ij}$
is the symmetric and traceless part of $-\delta F/\delta Q_{ij}$, 
and
is obtained from the Landau-de Gennes free energy~\cite{de1993physics}
\begin{align}
F&=\int f \mathrm{d}^2 r,
\label{eq:F}
\\
f&= 
\frac{A}{2}\mathrm{Tr} \, \mathbf{Q}^2
+ 
\frac{C}{4}\left(\mathrm{Tr} \, \mathbf{Q}^2\right)^2
+ 
\frac{K}{2}\left(\bs{\nabla}\mathbf{Q}\right)^2
-
\frac{1}{2}\bs{E}\cdot\mathbf{Q}\cdot\bs{E}.
\label{eq:f}
\end{align}
The first two terms of the free energy density control the magnitude of the scalar order parameter $S$.
Note that the term proportional to $\mathrm{Tr}\, \mathbf{Q}^3$ 
identically vanishes for the two-dimensional nematic order parameter.
The third term is the Frank elastic energy under the one-constant approximation,
and the fourth term describes the coupling to the quenched random field $\bm{E}$.
The molecular field is obtained as
\begin{align}
H_{xx} &= - \left(2 A + C S^2 \right) Q_{xx} + 2 K \nabla^2 Q_{xx} 
+ \frac{1}{2} \left(E_x^2 - E_y^2 \right).
\label{eq:Hxx}
\\
H_{xy} &= - \left(2 A + C S^2 \right) Q_{xy} + 2 K \nabla^2 Q_{xy} 
+  E_x E_y.
\label{eq:Hxy}
\end{align}
}
The stress tensor \Bdd{consists} of the passive stress 
\begin{equation}    
\bs{\upsigma}^\mathrm{e}=
-\lambda S \mathbf{H}+\mathbf{Q}\cdot\mathbf{H}-\mathbf{H}\cdot\mathbf{Q},
\label{eq:sigmae}
\end{equation}
and the active stress
\begin{equation}    
\bs{\upsigma}^\mathrm{a}=-\alpha\mathbf{Q}.
\label{eq:sigmaa}
\end{equation}
We assume an extensile active stress\Bdd{, so} the activity parameter $\alpha$ is positive.

We model the quenched random field \Bdd{as}
\begin{equation}
\bE = E_0 \be(\br) 
\label{eq:randomE}
\end{equation}
where $\be(\br) = (\cos \theta_e, \sin \theta_e)$ 
is a random unit vector 
\Bdd{where the} angle $\theta_e$ \Bdd{is} uniformly \Bdd{distributed}  in $[0, 2\pi)$.
\Add{In most of our numerical simulations, we
}
{implement the random field on a square lattice 
with the grid size $\Add{\Delta x}$,
and choose the angle $\theta_e$ randomly at each grid point.}
Therefore, it satisfies
\begin{align}
\langle \be(\br) \be(\br') \rangle = \frac12 \bI \, {\delta_{\br, \br'},}
\label{eq:correlationEgrid}
\end{align}
{
where $\delta_{\br, \br'}$ is the Kronecker delta \Bdd{defined on}
a lattice.
\Bdd{In the continuum limit, where the} 
typical \Bdd{length scale is} much larger than $\Add{\Delta x}$,
it is convenient to replace it with the expression
\begin{align}
\langle \be(\br) \be(\br') \rangle = \Add{\frac{(\Delta x)^2}{2}\bI} \delta(\br - \br').
\label{eq:correlationE}
\end{align}
\Delete{Note that both expressions give $\xi_e^2$ 
over the grid containing $\br=\br'$.}
The contributions of the \Delete{the} random field 
and Frank elasticity to the free energy are estimated as
$E_0^2 S_0$ and $KS_0^2/l_Q^2$, respectively,
where $S_0$ is the typical magnitude of the scalar order parameter
and $l_Q$ is the correlation length of $\bQ$.
The orientational correlation length \Bdd{decreases with
increasing disorder strength,}
and its lower bound is given by $l_Q \sim \Add{\Delta x}$.
Therefore, we define the effective disorder strength as 
\begin{align}
D_K = \frac{E_0^2 \Add{(\Delta x)}^2}{K S_0}.
\end{align}
On the other hand, the contributions to the molecular field by the random field 
and the active stress are estimated as $E_0^2$ and $\alpha$, respectively.
Thus we are led to the other definition of the dimensionless disorder strength,
\begin{align}
D_\alpha = \frac{E_0^2}{\alpha}.
\end{align}

{\it \label{sec:flowrandom} Disorder-dominated regime.} --
In the disorder\Bdd{-}dominated regime with $D_K \gg 1$ and $D_\alpha \gg1$,
the director \Bdd{aligns} with 
the local \Add{random field} and \Bdd{becomes} frozen.
In this case, we have approximately $\bn(\br) = \be(\br)$, $\bH = \bm{0}$ 
and $\sigmae = \bm{0}$ in the stationary state,
and the Frank elastic term in the molecular field is negligible.
Thus, from Eqs.(\ref{eq:Hxx}),(\ref{eq:Hxy}), 
the scalar order parameter satisfies
\begin{equation}
- S \left( 2A + C S^2 \right) + E_0^2 = 0,
\label{eq:S0}
\end{equation}
\Bdd{whose solution defines} $S_0$.
The nematic order parameter is given by
\begin{align}
\bQ(\br,t) &= S_0 \left[ \be(\br) \be(\br) - \frac12 \bI \right].
\label{eq:Qe}
\end{align}
Since the molecular field \Delete{is balanced and} 
vanishes in the steady state, 
the flow velocity is determined solely by the active stress.
For ${\rm Re} \ll 1$,
the velocity field is obtained by \Bdd{neglecting} 
the terms on the left hand side
of Eq.(\ref{eq:dvdt}) as
\begin{equation}
0 = \frac{1}{{\rm Re}} \nabla^2 \bv 
- \nabla p + \nabla \cdot \sigmaa.
\label{eq:vstatic}
\end{equation}
Solving (\ref{eq:vstatic}) under the incompressibility condition
and with (\ref{eq:sigmaa}),
we obtain the velocity field in the Fourier representation,
\begin{equation}
\bv^{\bm{k}}= -\alpha {\rm Re} \, \frac{\bI - \bkhat \bkhat}{k^2} \cdot 
\left( 
i\bk \cdot \bQ^{\bk}
\right),
\label{eq:vk}
\end{equation}
which \Bdd{yields} the velocity structure factor
\begin{align}
\Sigma_v(\bk) &= \left\langle \left| \bv^{\bk} \right|^2 \right\rangle
\nonumber\\
& 
= \frac{(\alpha {\rm Re})^2}{k^2}
\left(
\left\langle \left| \bkhat \cdot \bQ^{\bk} \right|^2 \right\rangle
-
\left\langle \left| \bkhat \cdot \bQ^{\bk} \cdot \bkhat \right|^2 \right\rangle
\right).
\label{eq:Svk}
\end{align}
The correlation function of the nematic order parameter 
\Bdd{is obtained from} 
from Eqs.(\ref{eq:correlationE},\ref{eq:Qe}) as
\begin{align}
\left\langle  Q_{ij}(\br) Q_{lm}(\br') \right\rangle
= \frac14 S_0^2 \Add{(\Delta x)}^2 \left( \delta_{il} \delta_{jm} + \delta_{im} \delta_{jl} \right) 
\delta(\br - \br'),
\label{eq:QQr}
\end{align}
and accordingly
\begin{align}
\left\langle  Q_{ij}^{\bk} Q_{lm}^{-\bk} \right\rangle
= \frac14 S_0^2 \Add{(\Delta x)}^2 \left( \delta_{il} \delta_{jm} + \delta_{im} \delta_{jl} \right).
\label{eq:QQk}
\end{align}
Substituting this into Eq.(\ref{eq:Svk}), we obtain
\begin{align}
\Sigma_v(\bk) = \frac{(\alpha {\rm Re} S_0 \Add{\Delta x})^2}{4 k^2}.
\label{eq:Sigmavk2}
\end{align}
The velocity correlation funcion 
in the real space is given by the inverse Fourier transform of $\Sigma_v(\bk)$
as
\begin{align}
\left\langle \bv(\br) \cdot \bv(\br') \right\rangle 
&= - \frac{ (\alpha {\rm Re} S_0 \Add{\Delta x})^2}{8\pi} \ln \left| \br - \br' \right|.
\label{eq:Cvr}
\end{align}

{\it \label{sec:results}Numerical simulation.} --
We solved Eqs.~(\ref{eq:dvdt},\ref{eq:dQdt}) numerically
on a square lattice with the fourth-order Runge-Kutta method.
\Bdd{We enforced the} incompressibility condition 
\Bdd{using} 
the simplified MAC method 
on a staggered lattice~\cite{amsden1970simplified}.
The main sublattice is used for the field variables $\mathbf{Q}$,
$p$, $\boldsymbol{\upsigma}$, $\mathbf{u}$, $\boldsymbol{\upomega}$ 
and $\mathbf{H}$,
and the other two sublattices are assigned to $v_x$ and $v_y$.
The calculation is performed on a $N_x \times N_y$ lattice 
with the grid size $\Delta x = \Delta y = 2$ 
and the step time increment $\Delta t=0.01$.
We \Bdd{imposed} periodic boundary conditions and
used \Bdd{the} Fast Fourier Transform 
to solve the \Bdd{Poisson} equation
for the pressure at each time step.
For the numerical analysis, we used the parameter values 
\begin{align}
A&=-0.16, \,
C=0.89, \,
K=1,
\label{eq:parameter1}
\\
\lambda&=0.1, \, \mathrm{Re}=0.1,\, \gamma=10. 
\label{eq:parameter2}
\end{align}
{
We varied the activity parameter $\alpha$ from $-0.2$ to $0.2$
to study both extensile and contractile systems.
}
\Add{In most of our simulations, the}
direction $\be(\br)$ of the random field \Bdd{was} 
randomly chosen at each grid point.
\Delete{which means $\xi_e=2$. }
The field strength is varied in the range $0 \le E_0 \le 0.7$.
\Add{We have also checked 
the effects of a spatially correlated random field,
which is prepared  by solving the diffusion equation for $\mathbf{E}$ starting from the grid-wise random field
and normalizing the amplitude to $E_0$.
The resulting random field has an approximately
Gaussian correlation function $C_E(\br-\br')=
\langle{\bE(\br)\cdot \bE(\br')\rangle}$.
We define 
the standard deviation of this distribution
as the disorder correlation length $\xi_E$.
We studied the cases $\xi_E=4.0, 6.0, 8.0$,
in addition to the grid-wise random field,
which gives $\xi_E = 2(1-e^{-1/2}) \approx 0.8$.
}
The scalar order parameter in the passive ($\alpha=0$) and stationary system
is obtained from Eq.~(\ref{eq:S0}) 
as $S_0 \simeq 0.60$ for $E_0=0$ and
{increases with $E_0$. 
We confirmed that $S$ does not \Bdd{exceed} unity 
for the strongest random field studied ($E_0=0.7$).}
The defect core radius is given by
{$\xi = \sqrt{K/|A|} \simeq 2.5$}.
The balance between the activity and Frank elasticity defines the \Bdd{length scale}
$l_\alpha = \sqrt{K/{|\alpha|}}$
{, which \Bdd{gives}
$l_\alpha \simeq 2.2$ for $|\alpha| = 0.2$ and 
$l_\alpha \simeq 3.2$ for $|\alpha| = 0.1$.
}
\Add{Thus we have ensured that $l_\alpha$ exceeds the grid size.}
The system size is fixed to 
$N_x=N_y= {256}$ 
so that $L=N_x \Delta x = N_y \Delta y = {512}$.
For the initial conditions,
we \Bdd{initialized} the velocity to zero and \Bdd{imposed}
small random fluctuations
around zero for $\mathbf{Q}(\bm{r},0)$, 
assuming a quench from the isotropic quiescent state.
\Bdd{Specifically,} the scalar order parameter 
and the director angle at each grid point are randomly chosen 
in the \Bdd{intervals} $[0,0.1]$ and $[0,2\pi]$, respectively.
We observed the total kinetic energy as a function of time
to confirm that the system reached \Bdd{a} dynamical steady state\Delete{s},
typically by $t = 10000$ \Bdd{in} active turbulence \Bdd{regimes}.
We calculated the data over the time window 
{$20000 < t \le 40000$}
with the time interval $t_0 = 100$, 
and took the ensemble average over {8}
independent \Bdd{realizations.}

\begin{figure}[thb]
\includegraphics[scale=1.0]{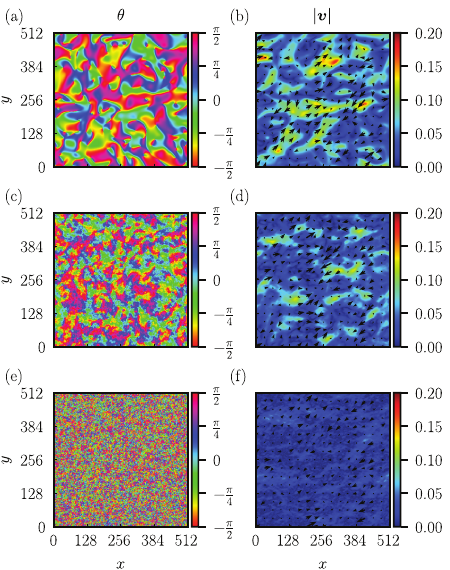}
\caption{\label{fig:snapshot} 
Snapshots of the director angle $\theta(x,y)$ \Bdd{(first column)} 
and \Bdd{the} velocity \Bdd{field} $\boldsymbol{v}(x,y)$ 
\Bdd{(second column), where the color indicates the} magnitude 
\Bdd{and} black arrows \Bdd{represent} the vector field.
\Bdd{The activity is fixed at} $\alpha =0.2$\Bdd{,} and
the \Bdd{random field strengths are} 
(a)(b) $E_0=0$, (c)(d) $E_0=0.4$, and (e)(f) $E_0=0.7$. 
}
\end{figure}
\begin{figure}[thb]
\includegraphics[scale=1.0]{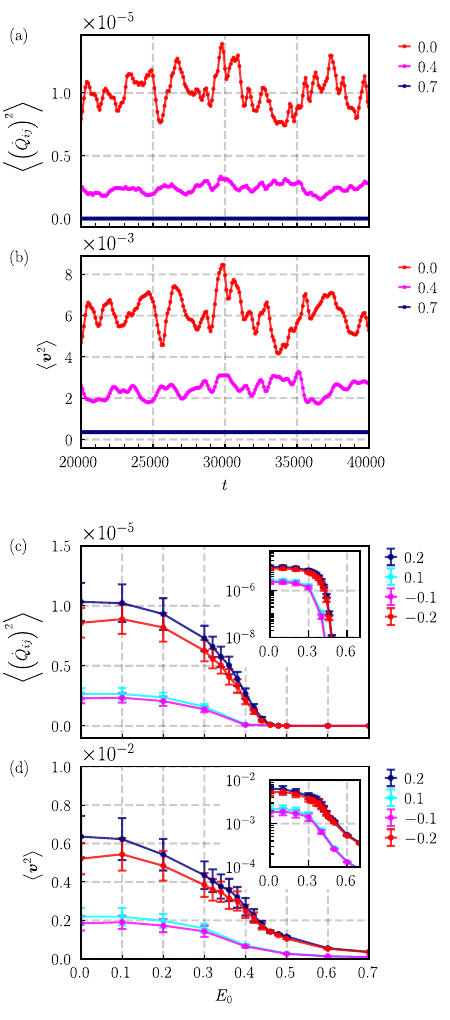}
\vspace{0mm}
\caption{\label{fig:meansquare} 
{
Mean square of the \Bdd{time derivative} of $Q_{ij}$
and \Bdd{the} flow velocity $\bv$ {for $\alpha=0.2$}.
(a), (b)\Bdd{:} Time evolution for a single sample, \Bdd{at} $E_0=0.0, 0.4, 0.7$.
(c), (d)\Bdd{:} Averages over time and 8 samples\Bdd{,}
plotted \Bdd{as functions of} the \Bdd{random} field strength $E_0$ 
\Delete{and} 
for $\alpha=0.2, 0.1, -0.1, -0.2$.
Error bars \Bdd{represent} the standard deviation.}
Insets: semi-\Bdd{logarithmic} plots.
}
\end{figure}

{\it \label{subsec:dynamical} Spatial patterns and orientational freezing.} --
In Fig.~\ref{fig:snapshot}, we show the snapshots of the director angle $\theta(\bm{r}, t)$ 
and the vorticity $\omega(\bm{r},t)$ in the dynamical steady states
for $\alpha =0.2$. 
For $E_0=0$, 
active turbulence containing topological defects and vortices \Bdd{is} reproduced 
[Fig.\ref{fig:snapshot}(a)(b)].
For $E_0=0.4$, 
the director pattern becomes \Bdd{jagged} while maintaining the characteristic large-scale 
structure of nematic defects. 
The velocity field is smoother but \Add{streamlines with 
wavy shapes} appear
[Fig.\ref{fig:snapshot}(c)(d)].
For $E=0.7$, 
the director orientation becomes completely random, and the flow pattern obtains
\Add{streak-like structures of various sizes} 
[Fig.\ref{fig:snapshot}(e)(f)].

The dynamics slow\Delete{s} down as 
we increase the field strength.
In Fig.~\ref{fig:meansquare}, we show the mean square of 
the time \Delete{-} derivative 
of the order parameter $\dot{Q}_{ij}$ and flow velocity.
{
The time evolution of these quantities for $20000 < t \le 40000$ is shown 
in Fig.~\ref{fig:meansquare}(a),(b), 
for $\alpha = 0.2$ fixed and $E_0 = 0, 0.4$ and $0.7$. 
\Bdd{These quantities exhibit} 
large fluctuations around their mean values\Bdd{,} 
with a typical timescale $\sim 10^3$ for $E_0=0.0$. 
Both the mean values 
and fluctuations get smaller as $E_0$ is increased. For $E_0=0.7$,
$\langle{\dot{Q}_{ij}^2\rangle}$ 
vanishes\Bdd{,} 
while the velocity \Bdd{maintaines} a small \Bdd{but finite} 
magnitude.
These data justify \Bdd{using statistical averages} 
over the same time window
to be shown in the rest of this paper. 
In Fig.~\ref{fig:meansquare}(c),(d), we show the time-averaged values
for $\alpha=0.2$, $0.1$, $-0.1$ and $-0.2$ as functions of $E_0$.
We find that the extensile  ($\alpha>0$) and contractile ($\alpha<0$) cases
\Bdd{show} qualitatively \Bdd{similar behavior}.
Quantitatively, the extensile \Bdd{systems exhibit
values approximately}  $10-20\%$ larger 
than the contractile cases with the same magnitude of $|\alpha|$,
for both $\dot{Q}_{ij}^2$ and $v^2$ and for $E_0 \le 0.4$.
For $E_0 \ge 0.5$, the differences between them become very small.
The ratio\Bdd{s} of these quantities between $\alpha = 0.2$ and $\alpha=0.1$ are 
about 4 for $\dot{Q}_{ij}^2$ and 3 for $v^2$ at $E_0=0$.
For $\alpha=0.2$,  $\dot{Q}_{ij}^2$ shows a sharp drop \Delete{to} below $10^{-6}$
at $E_0=0.44$, while a similar drop \Bdd{occurs} at a smaller 
\Delete{value of} $E_0$ for $\alpha=0.1$;
see the semi-log plots in the inset of Fig.~2(c).
\Bdd{We henceforth} focus on the extensile case $\alpha=0.2$.}
The director dynamics completely freeze\Delete{s} 
at {$E_0\approx 0.5$}.
On the other hand, the mean square velocity decreases only gradually 
as we increase the \Bdd{disorder strength}.
{Its} decay \Bdd{for $E_0> 0.44$}  is {slower} 
than that of the {orientatio\Add{n}al order parameter.}
For $E_0=0.7$, the strongest field we studied,
the mean square velocity still remains at 
\Bdd{approximately 6 \%}  of its value at $E_0=0$.
Note that the effective disorder strengths are  $D_K \simeq 0.97$ 
and $D_\alpha \simeq 0.96$ for $E_0=0.44$, which are both close to unity.
Therefore, it would be reasonable 
to \Bdd{distinguish} the medium and strong disorder regimes 
at this value of $E_0$.
\Add{
We also examined the dynamics under
a spatially correlated random field.
As we increase $\xi_E$ from $0.8$ to $8.0$
with the fixed amplitude $E_0=0.4$,
$\langle \dot{Q}_{ij}^2 \rangle$ 
decreases from 
$2.6\times 10^{-6}$ to $7.0\times10^{-12}$,
indicating that
the director dynamics depends 
significantly
on the random-field correlation length.
On the other hand, $\langle{v^2 \rangle}$ 
shows
only minor decrease from $2.7 \times 10^{-3}$
to $1.9 \times 10^{-3}$.
Thus, the remnant flow in the orientationally 
frozen state is robust and only weakly 
dependent on the structure of the random field.
}
\\

\begin{figure}[htbp]
\includegraphics[scale=1.0]{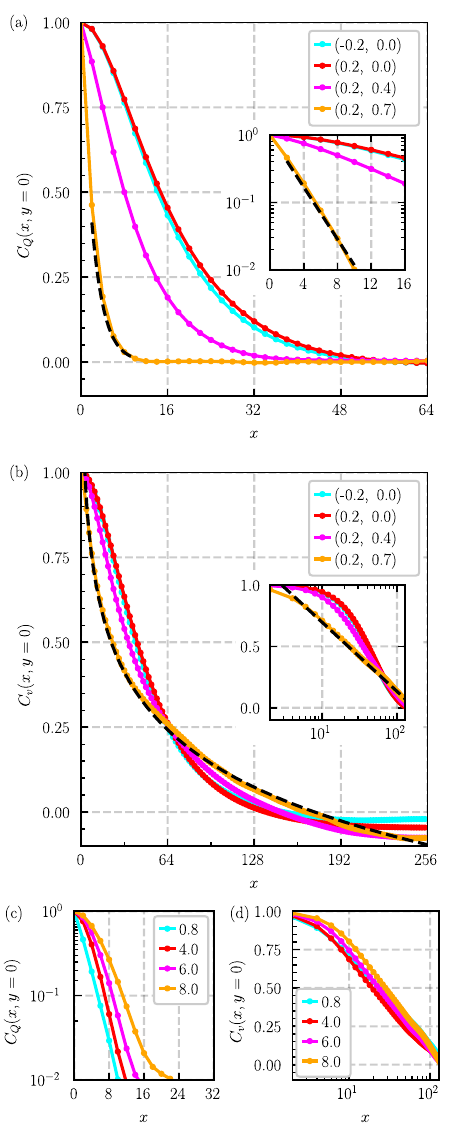}
\vspace{0mm}
\caption{\label{fig:corrfunc} 
\Bdd{Orientational and velocity correlation functions.}  
\Bdd{(a), (b):} 
$(\alpha, E_0) = (-0.2,\ 0.0),\ (0.2,\ 0.0),\ (0.2,\ 0.4),\ (0.2,\ 0.7)$ 
\Add{with $\xi_E = 0.8$.
(c), (d): 
$(\alpha, E_0) = (0.2,\ 0.7)$ with $\xi_E = 0.8,\ 2.0,\ 4.0,\ 8.0$.  
}
\Bdd{(a): $C_Q(r)$ (inset: semilog-$y$ plot).}  
\Bdd{Dashed} lines \Bdd{indicate} the \Add{fitting} function 
$\exp(-r/r_Q)$ with $r_Q = 2.3$.  
\Bdd{(b): $C_v(r)$} \Add{ (inset: semilog-$x$ plot).}  
\Bdd{Dashed} line \Bdd{indicates} the \Add{fitting} function 
$-A \ln(r/r_v)$ with $A = 0.25$ and $r_v = 173$.  
\Add{
(c): $C_Q(r)$ (semilog-$y$ plot).  
(d): $C_v(r)$ (semilog-$x$ plot).}
}
\end{figure}

{\it \label{sec:dist} Correlation functions and correlation lengths.} --
In Fig.~\ref{fig:corrfunc}, we show the spatial correlation functions 
for the {orientational order parameter} and flow velocity, which are defined by
\begin{equation}
\Add{
C_Q(\bs{r})=
\frac
{\langle \bs{Q}(\bs{r}+\bs{r'},t) : \bs{Q}(\bs{r'},t) \rangle}
{\langle \bs{Q}(\bs{r'},t)^2 \rangle}
}
\end{equation}
and
\begin{equation}
\Add{
C_v(\bs{r})=
\frac
{\langle \bs{v}(\bs{r}+\bs{r'},t) \cdot \bs{v}(\bs{r'},t) \rangle}
{\langle \bs{v}(\bs{r'},t)^2 \rangle},
}
\end{equation}
respectively, where \Add{$\langle \cdots \rangle$} \Bdd{denotes}  
averages over $\bs{r'}$ and $t$,
and {8} independent \Bdd{realizations}.
By symmetry, the correlation functions \Bdd{depend only on
the distance $r$} .
We show their profiles along the $x$-axis in Fig.~\ref{fig:corrfunc}.
{
\Bdd{We again confirm} that the extensile case ($\alpha=0.2$)
and contractile case ($\alpha=-0.2$) show similar behaviors for $E_0=0.0$,
we \Bdd{henceforth} focus on the extensile case in the following,
and vary the field strength as $E_0=0.0$, $0.4$, and $0.7$.
}
The orientational correlation function $C_Q(r)$ 
\Bdd{decreases} monotonically with \Delete{a} positive curvature
 [Fig.~\ref{fig:corrfunc}(a)].
For $E_0=0.7$, it is \Bdd{well} fitted by the exponentional function
{$C_Q(r) = \exp(-r/r_Q)$ with $r_Q = \Add{2.3}$. } 
\Add{The fitting range is $2 \le r \le 10$.}
The semi-logarithmic plot in the inset of Fig.~\ref{fig:corrfunc}(a)
shows that the exponential decay also holds 
for the medium disorder case $E_0=0.4$, up to $r \approx 30$,
{while the decay is faster than exponential for $E_0=0$
as seen from the \Bdd{upward-convex shape} in the semi-log plot.
}
{Furthermore, for $E_0=0.7$, we confirmed that 
the orientational correlation functions for $-0.2 \le \alpha \le 0.2$ 
closely match \Delete{with} each other; 
the data are not shown as they 
are \Bdd{in}distinguishable at the scale of Fig.~\ref{fig:corrfunc}(a).}

\begin{figure}[htbp]
\includegraphics[scale=1.0]{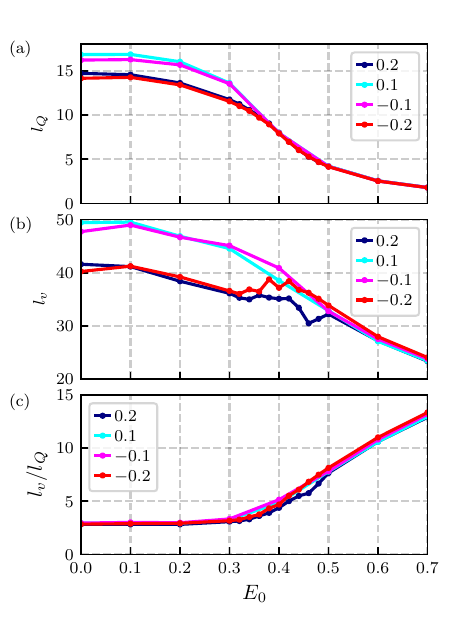}
\vspace{0mm}
\caption{\label{fig:corrlen} 
(a) \Bdd{Orientational} correlation length $l_Q$\Bdd{,}  
(b) velocity correlation length $l_v$\Bdd{, and} 
(c) \Bdd{the} ratio $l_v/l_Q$ 
as functions of {the \Bdd{random} field strength} $E_0$.  
}
\end{figure}

The velocity correlation function decays more slowly than the orientational one, 
and turns negative at 
{$r \sim L/4$ for $E_0=0$ and at $r \sim L/3$ for $E_0=0.7$} [Fig.~\ref{fig:corrfunc}(b)].
It \Bdd{decreases} monotonically up to $r = 0.5L$ 
\Bdd{and is expected to} converge to zero for $r\to \infty$.
The function has a negative curvature in a narrow range near $r=0$,
which becomes narrower for a larger field strength.
For $E_0=0.7$, it is \Bdd{well} fitted by the logarithmic function $C_v(r) = - A \ln(r/r_v)$ \Add{over the range $2\le r <L/2$, 
yielding}
$A=\Add{0.24}$ and $r_v=\Add{173}$.
\Delete{in the range $2 < r< r_v$. }
{In addition, for \Add{a} $L=256$ system, 
we found that the same fitting gives $r_v = 78$, 
\Bdd{suggesting} that $r_v$ is determined by\Bdd{,} 
and proportional to\Bdd{, the} system size.
}

\Add{To investigate the effects of a spatially correlated random field,
we show the orientational and velocity correlation functions 
for $\xi_E=0.8, 4.0, 6.0, 8.0$ in Fig.~\ref{fig:corrfunc}(c)(d),
where $\alpha=0.2$ and $E_0 = 0.7$ are fixed.
We find that $C_Q(r)$ decays approximatrely 
exponentially for $r>\xi_E$,
while it shows a slower decay for $r<\xi_E$ and at large distances 
where $C_Q(r) \ll 0.1$. On the other hand, $C_v(r)$ 
in the semilog-$x$ plot shows a constant slope over a wide range 
($8 < r < 50$) for all cases, while the decay is slower near the origin.
These results confirm the robustness of the exponential and
logarithmic decay of $C_Q(r)$ and $C_v(r)$, respectively. 
}

We define the correlation lengths {$l_Q$} and $l_v$ 
by {$C_Q(l_Q)$}$=1/2$ and $C_v(l_v)=1/2$, respectively.
They are plotted in Fig.~\ref{fig:corrlen}(a){(b)} as functions of the field strength.
The orientational correlation length shows a rapid decay
between $E_0=0.2$ and $0.5$.
The decay of 
the velocity correlation length is slower{, and \Bdd{exhibits} 
large fluctuations in the transition region $0.3<E_0 < 0.5$\Bdd{,
reflecting} sample dependence.}
The ratio $l_v/${$l_Q$}\Bdd{, as shown} in Fig.~\ref{fig:corrlen}{(c)},
increases to {$13$} 
at $E_0=0.7$ from {2.8} at $E_0=0$.
\\

{\it \label{sec:discuss} Discussion.} --
{
We found a marked contrast between the orientational and flow properties, 
which \Bdd{can be} summarized as follows: (i) \Bdd{The} 
dynamics of the orientational order parameter
\Bdd{freezes} completely at and above a critical random field strength, 
while the flow velocity only gradually decreases \Bdd{with increased randomness} and remains finite \Bdd{even} 
in the strong disorder limit.
(ii) The spatial correlation of the orientational order parameter is short-ranged 
and exhibits exponential decay for strong disorder, while the flow velocity \Bdd{retains} long-range 
correlation\Bdd{s, characterized by} 
logarithmic decay for strong disorder.
We have also compared the extensile and contractile systems, and found only 
\Bdd{minor differences} between the two \Bdd{when} 
the \Add{magni}tude of activity $|\alpha|$ is the same.
The difference becomes negligible \Bdd{in the} strong disorder \Bdd{regime}.
}

The exponential decay of the orientational correlation function $C_Q(r)$ in the strong disorder regime\Delete{s}
[Fig.~\ref{fig:corrfunc}(a)] 
\Bdd{agrees}
with the previous result on 2D random-field nematics~\cite{yu1998exponential}.
This suggests that the director texture is determined 
by \Bdd{the} balance between the random field and Frank elasticity,
and is \Bdd{only weakly} affected by the active flow.
{This is supported by the fact that $C_Q(r)$ for $-0.2 \le \alpha \le 0.2$ are \Add{nearly}
indistinguishable from each other for $E_0=0.7$.} 
Earlier studies on the random-field XY model \Bdd{reported}
correlation \Bdd{decaying} 
faster than exponential in two dimensions~\cite{dieny1990xy, gingras1996topological}.
The difference is attributed to the \Bdd{difference in the} 
symmetry of \Bdd{the} random anisotropy 
in nematics and the XY model~\cite{yu1998exponential}. 
Without the randomness, 
the orientational correlation function decays faster than 
\Bdd{exponentially} as seen in Fig.~\ref{fig:corrfunc}(a).
In this case, the correlation functions are characterized by the vortex size~\cite{giomi2015geometry},
and the velocity correlation function $C_v(r)$ has a negative curvature at short distances,
as seen in Fig.~\ref{fig:corrfunc}(b)\Bdd{, which agrees}
with the analytical result~\cite{giomi2015geometry}.
As we increase the random field, the range with a negative curvature shrinks and  
the velocity correlation function is better approximated by the logarithmic function. 
This is in agreement with the 
analytical result for the strong disorder [Eq.(\ref{eq:Cvr})],
and reflects the structure of the Green function of the Stokes equation in two dimensions.
In three dimensions, the velocity correlation function should decay as $1/r$ in the strong disorder limit. 
\Add{Note that we cannot fully reach the strong disorder limit
in our numerical simulations.
The} effective disorder strengths are $D_K \simeq 2.0$ and 
$D_\alpha \simeq 2.5$ for $E_0=0.7$ {and $\alpha=0.2$}.
\Add{In addition,} 
the orientational correlation length $r_Q=\Add{2.3}$
\Add{for $E_0=0.7$} is close to the \Add{grid size $\Delta x$}.
\Add{The amplitude $E_0$ was not increased further 
in order to ensure that the scalar order parameter remained 
within a physically reasonable range 
and that the correlation length stayed larger than the grid size.
}
\Add{Still, the logarithmic} function for the velocity correlation \Add{fits}
the numerical results \Add{well,} with errors within $\pm 6$\% 
over the range $4 \le r\le 110$, where $C_v(r) \ge 0.1$.
The error increases to 16\% at $r=\Add{\Delta x}=2$
due the discreteness of the lattice. 
These \Bdd{findings} confirm the validity of 
the analytical results obtained 
by the continuum approximation over a wide distance range. 
} 

The dependence\Delete{s} of the orientational and velocity correlation lengths 
on the field strength \Bdd{is} also in marked contrast.
For weak disorder, both lengths are \Bdd{proportional} 
to the vortex size and the ratio $l_v/l_Q$ is small.
For strong disorder, the velocity correlation decays {monotonically}
even when the {orientational} correlation length 
\Bdd{approaches a} constant, and 
thus $l_v/l_Q$ diverges in the strong disorder limit.
The slowing down of the increase of $l_v/l_Q$ in the strong disorder regime
 [Fig.~\ref{fig:corrlen}(b)]
is explained by \Bdd{the fact} 
that \Delete{the} {$l_Q$} has a lower bound determined by
the defect core size{, and that $l_v$ has \Bdd{an} 
upper bound determined by the system size.
(Note that $l_v$ cannot exceed $L/2$ due to the periodic boundary condition.)}
\Add{Since the Stokes equation lacks a characteristic length scale, 
it is natural that $l_v$ is proportional to the system size.
}

Finally, we consider the competition between the active flow and \Bdd{the} random field.
The flow-aligning effect on the nematic order parameter is represented by the term 
$\lambda S \mathbf{u}$ in Eq.~(\ref{eq:dQdt}),  
the magnitude of which is estimated as 
$\lambda S_0 v_{\rm rms} /l_v$. 
The contribution of the random field \Bdd{to} the term $\gamma^{-1} \mathbf{H}$
is estimated as $\gamma^{-1} E_0^2 S_0$.
Averaging it over the area $l_Q^2$ of an orientationally correlated region,
which contains $N \sim (l_Q/\Add{\Delta x})^2$ sites,
we get $\gamma^{-1} E_0^2 S_0 /\sqrt{N} 
\sim \gamma^{-1} E_0^2 S_0 \Add{\Delta x}/ l_Q$.
Thus the ratio between the flow-aligning and random-field terms is
estimated as
${\gamma \lambda v_{\rm rms} l_Q}/(E_0^2 l_v \Add{\Delta x})$.
In our simulation, 
this ratio becomes {$0.33$} for $E_0=0.2$ {and $\alpha=0.2$}, 
which \Bdd{supports} 
the observation that the active flow has a minor effect 
in the medium ($0.2 < E_0 < 0.44$) and strong ($E_0 > 0.44$) disorder regimes.
It is also consistent {with} the fact that the mean flow velocity and correlation lengths start to decrease
around $E_0 = 0.2$.
\Add{
Note that the dependence of this ratio on the activity parameter and the random field strength arises mainly from the
$v_{\rm rms}$ and $1/E_0^2$ and that the former is proportional to $\alpha$ 
in the weak disorder regime. 
The other factor $l_Q/l_v$ also depends on $\alpha$ and $E_0$ but more weakly:
$l_Q \sim l_\alpha \propto |\alpha|^{-1/2}$ in the weak disorder regime,
and $l_Q/l_v$ changes roughly linearly with $E_0$ in the strong disorder regime [Fig.~4(c)].
Therefore, the main differences in the system's behavior 
due to the activity and disorder strength 
can be attributed to the active stress ($\propto \alpha$) 
and the random field free energy ($\propto E_0^2$), 
which enter the dynamical equations. 
}

In summary, quenched disorder introduces \Bdd{distinctive effects}into the physics of active nematics.
\Bdd{In the} strong disorder \Bdd{regime}, the director texture 
\Bdd{becomes} frozen and \Bdd{is governed}  
by the balance between
the randomness and Frank elasticity, while \Bdd{the} 
active flow \Bdd{persists} with long-range \Bdd{correlations} and facilitates material transport.
We hope that the present work \Bdd{will stimulate} experimental studies on the flow properties 
of cellular and subcellular systems with orientational order.
\\

{\it Acknowledgements.} --
Y.K. acknowledges support of the work through the International Joint Graduate Program in
Materials Science at Tohoku University and JST SPRING, Grant Number JPMJSP2114.




%

\end{document}